\documentclass[twocolumn,showpacs,pra,aps,amsmath,amssymb,superscriptaddress]{revtex4}

\usepackage{graphicx}
\usepackage{bm}
\usepackage{color}

\begin{document}

\title{
Long distance entanglement in one-dimensional quantum systems under sinusoidal deformation
}
\author{Toshiya Hikihara}
\affiliation{Faculty of Engineering, Gunma University, Kiryu, Gunma 376-8515, Japan}
\author{Takafumi Suzuki}
\affiliation{
Research Center for Nano-Micro Structure Science and Engineering, Graduate School of Engineering, University of Hyogo, Himeji, Hyogo 671-2280, Japan
}

\date{\today}

\begin{abstract}
We investigate entanglement generation in one-dimensional quantum spin systems with the sinusoidal deformation. In the system, the energy scale of each local term in the Hamiltonian is modified according to a position-dependent function $\sin^\alpha\left[\frac{\pi}{N} \left( x - \frac{1}{2} \right) \right]$, where $x$ is the position of the local term and $N$ is the length of the system. We show that at zero temperature the system with $\alpha \ge 2$ is able to generate a sizable entanglement between two spins at open edges even when the two spins are infinitely far apart. This long-distance entanglement is rather robust against thermal fluctuations and survives up to a temperature that decays with the system size slowly, in an algebraic form. 

\end{abstract}

\pacs{
03.67.Bg,
03.67.Hk,
75.10.Pq
}

\maketitle

\section{Introduction}\label{sec:intro}

Entanglement is an essential resource in quantum information tasks.\cite{NielsenC}
In particular, the generation of a large entanglement between two parties {\it located far away from each other} is a crucial ingredient in quantum information processing such as quantum computation and quantum teleportation.
One possible way to create an entanglement between distant parties is to connect them by a one-dimensional (1D) quantum many-body system including quantum spin chains and correlated fermion/boson systems.
Such a quantum system which is able to sustain a sizable entanglement between qubits at a distance has been sought under the concept of long distance entanglement (LDE).\cite{VenutiBR2006,VenutiBR2007,VenutiGIZ2007,LiSCSS2005,FerreiraS2008,XiL2008,FerreiraLS2010,ReslenB2009,WojcikLKGGB2005,HartmannRP2006,WichterichB2009,GalveZKLH2009,SodanoBB2010}

The realization of LDE is a challenging task since entanglement in a system with short-range interactions usually decays quite rapidly.
However, the considerable efforts in recent years have revealed that there are some systems which can generate LDE.
A typical example of the systems exhibiting LDE is the spin-1/2 bond-alternating chain with two additional spins weakly coupled to the open ends.\cite{VenutiBR2006,VenutiGIZ2007}
The ground state of the system consists of entangled pairs of spins (singlet pairs, in the words of quantum magnetism) in the bulk and two effectively-free spins at the open edges.
At a sufficiently low temperature, the edge spins form a singlet pair across the bulk spins and realize LDE.
A similar mechanism applies also to other systems with gapful excitations.\cite{VenutiBR2006,LiSCSS2005,FerreiraS2008,XiL2008}
Setups utilizing 1D critical and 2D spin systems\cite{VenutiBR2006,VenutiBR2007,VenutiGIZ2007,FerreiraLS2010} and bosons in a 1D optical lattice\cite{ReslenB2009} to mediate LDE between edge qubits as well as dynamical setups to generate LDE\cite{ReslenB2009,WojcikLKGGB2005,HartmannRP2006,WichterichB2009,GalveZKLH2009,SodanoBB2010} have also been proposed.

\begin{figure}
\begin{center}
\includegraphics[width=75mm]{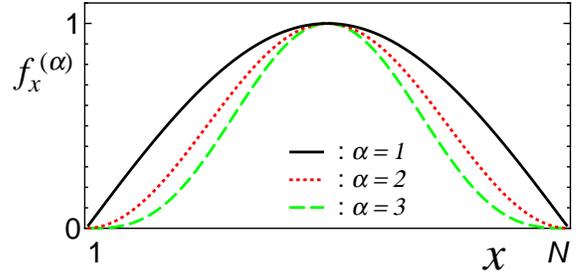}
\caption{(Color online) 
Rescaling function $f_x^{(\alpha)}$ of the sinusoidal deformation, Eq.\ (\ref{eq:fx}) with $\alpha = 1, 2,$ and $3$.
$f_x^{(\alpha)}$ is maximum at the center of the system, $x = (N+1)/2$, and becomes zero at open edges, $x = 1/2, N+1/2$.
}
\label{fig:scaling_function}
\end{center}
\end{figure}

In this paper, we propose a new class of 1D systems which exhibit LDE, that is, the systems under the sinusoidal deformation (SD).\cite{GendiarKN2009,HikiharaN2011,Katsura2011,MaruyamaKH2011,GendiarDLN2011,ShibataH2011,Katsura2011B}
The SD is introduced as follows.
First, we consider a Hamiltonian consisting of local terms,
\begin{eqnarray}
\mathcal{H}_0 = \sum_x h_x,
\end{eqnarray}
where $x$ denotes the center position of the local Hamiltonian. 
For instance, $x=(l+l')/2$ for the exchange term between the spins at $l$th and $l'$th sites, $h_x = {\bf S}_l \cdot {\bf S}_{l'}$.
Then, in the system under SD, the energy scale of the local Hamiltonian at the position $x$ is modified according to a rescaling function;
Namely,
\begin{eqnarray}
\mathcal{H}_{\rm SD} = \sum_x f_x^{(\alpha)} h_x,
\end{eqnarray}
with
\begin{eqnarray}
f_x^{(\alpha)} 
= \sin^\alpha\left[\frac{\pi}{N} \left( x - \frac{1}{2} \right) \right],
\label{eq:fx}
\end{eqnarray}
where $N$ is the number of sites in the system.
The energy scale is a maximum at the center of the system, $x=(N+1)/2$, decreases smoothly as the position $x$ goes away from the center, and becomes zero at the open edges, $x=1/2, N+1/2$ (see Fig.\ \ref{fig:scaling_function}).

Employing an analytic argument as well as numerical methods, we study the entanglement between two spins at the open ends of the systems under SD.
We show that in the ground state of the systems with $\alpha \ge 2$ the entanglement between the edge spins remains finite even at the limit of infinitely-long system, that is, LDE is realized.
In particular, for $\alpha = 2$ the emergence of LDE can be proven exactly.
The amount of LDE in the ground state increases with $\alpha$.
We also examine how robustly the LDE survives at finite temperatures.
It is found that the LDE becomes more fragile against thermal fluctuations as $\alpha$ is larger; 
In the $N$-site system with $\alpha \ge 2$, the temperature at which the entanglement between edge spins vanishes decreases with $N$ in the scaling form $\sim N^{-\alpha}$.
Our results provide materials for finding the optimal $\alpha$ to maximize the amount of LDE at the lowest temperature one can reach.

The paper is organized as follows.
The definition of models considered in this paper is presented in Sec.\ \ref{subsec:model}.
In Sec. \ref{subsec:SSD_exact}, we review known results for the systems under SD with $\alpha=2$ and show that LDE between edge spins is realized in the ground state of the systems.
We then present our numerical results for the ground state and finite temperatures in Sec.\ \ref{subsec:groundstate} and \ref{subsec:finiteT}, respectively.
Section\ \ref{sec:Conclusion} contains the summary and concluding remarks.

\section{Sinusoidal deformation}\label{sec:SD}

\subsection{Model Hamiltonian}\label{subsec:model}
We study the spin-1/2 antiferromagnetic XXZ chain under the sinusoidal deformation.\cite{GendiarKN2009,HikiharaN2011,Katsura2011,MaruyamaKH2011,GendiarDLN2011,ShibataH2011,Katsura2011B}
The model Hamiltonian is given by
\begin{eqnarray}
\mathcal{H} 
&=& J \sum_{l=1}^{N-1} f^{(\alpha)}_{l+1/2} 
\left( S^x_l S^x_{l+1} + S^y_l S^y_{l+1} + \Delta S^z_l S^z_{l+1} \right),
\label{eq:Ham}
\end{eqnarray}
where ${\bf S}_l = ( S^x_l, S^y_l, S^z_l )$ is the spin-1/2 operator at $l$th site and $\Delta$ is the parameter of the exchange anisotropy.
The rescaling function $f^{(\alpha)}_x$ is defined in Eq.\ (\ref{eq:fx}).
In this paper, we consider the cases of the Heisenberg chain ($\Delta = 1$) and the XY chain ($\Delta = 0$).
Note that the spin-1/2 XY chain is mapped to the 1D system of free spinless fermions by Jordan-Wigner transformation.

\subsection{Long-Distance Entanglement by Sine-Square Deformation}\label{subsec:LDE_in_SSD}\label{subsec:SSD_exact}

The SD was originally introduced in Ref.\ \onlinecite{GendiarKN2009} as an improved version of smooth boundary condition\cite{VekicW1993,VekicW1996} to suppress open boundary effects.
In particular, for the SD with $\alpha=2$, which is called the sine-square deformation, an interesting phenomenon was found: Despite of the presence of the open boundaries, the boundary oscillations in local quantities were suppressed completely.\cite{GendiarKN2009}
Subsequently, it was found that a 1D critical (gapless) system with the sine-square deformation had the ground state which was identical in the level of the wave function to the one of the same uniform system under the periodic boundary condition.\cite{HikiharaN2011,SSD_to_antiperiodic}
The equivalence of the ground-state wave functions was proven rigorously for several systems including 1D free-fermion system [equivalent to the XY chain, Eq.\ (\ref{eq:Ham}) with $\Delta=0$], the transverse-field Ising model at criticality, and the Gaussian model of the $c=1$ conformal-field theory\cite{Katsura2011,MaruyamaKH2011,Katsura2011B}.
Furthermore, it was found numerically that the same phenomenon occurred in a wide variety of 1D models with gapless excitations such as the spin-1/2 XXZ chain and two-leg ladder in magnetic field\cite{HikiharaN2011} and the (extended) Hubbard chain in a metallic phase\cite{GendiarDLN2011}.
These results suggest that the equivalence of the ground state between systems under the periodic boundary condition and the sine-square deformation is a generic feature of 1D critical systems belonging to the universality class of Tomonaga-Luttinger liquid.

In relation to LDE, the above observation immediately leads to an intriguing conclusion:
In the ground state of the system with the sine-square deformation, the spins at the open edges, which are at a distance of the length of the system, behave in completely the same way as the nearest-neighboring spins in a periodic system.
Since the entanglement between the neighboring spins in the periodic chain remains finite even in the thermodynamic limit, it follows that the edge spins in an open chain with the sine-square deformation also sustain the same finite amount of entanglement.
It is thereby concluded that the system under the sine-square deformation exhibits the LDE.
We will show later that the model (\ref{eq:Ham}) for $\alpha \ge 2$ also realizes the LDE in the ground state.

\section{numerical results}\label{sec:numerical}

In this section, we present our numerical results of the entanglement between edge spins in the spin chain under the sinusoidal deformation.
To evaluate the entanglement between two end spins, we calculate the concurrence defined as follows.\cite{HillW1997,Wootters1998}

We start from the density matrix (DM) for the whole system, i.e., 
$\rho = | \Phi_0 \rangle \langle \Phi_0 |$
at $T=0$ ($|\Phi_0 \rangle$ is the ground state) and 
$\rho = \frac{1}{Z} \exp\left( - \frac{\mathcal{H}}{k_B T} \right)$ 
for finite temperatures $T>0$ ($Z$ is the partition function and $k_B$ is the Boltzmann constant).
The reduced DM for the edge spins ${\bf S}_1$ and ${\bf S}_N$ is obtained as
$\rho_{1N} = {\rm Tr}_{\overline{1N}} \rho$, 
where ${\rm Tr}_{\overline{1N}}$ denotes the trace for the bulk spins at $l=2,...,N-1$.
The concurrence for the edge spins is then defined as
\begin{eqnarray}
\mathcal{C}(\rho_{1N}) = \max(0, \lambda_1-\lambda_2-\lambda_3-\lambda_4)
\label{eq:Conc}
\end{eqnarray}
where $\lambda_i$ ($\lambda_1 \ge \lambda_2 \ge \lambda_3 \ge \lambda_4$) are the square roots of the eigenvalues of the non-Hermitian matrix $\rho_{1N} \tilde{\rho}_{1N}$.
Here, $\tilde{\rho}_{1N}$, called the spin-flipped state, is defined by
\begin{eqnarray}
\tilde{\rho}_{1N} = (\sigma_1^y \otimes \sigma_N^y ) \rho_{1N}^{*}
(\sigma_1^y \otimes \sigma_N^y ),
\end{eqnarray}
where $\sigma^y$ is the $y$ component of the Pauli matrix.
The concurrence is equal to $1$ for the maximally entangled state while it is $0$ for a separable state.

Using relations between the matrix elements of the reduced DM $\rho_{1N}$ and expectation values of edge-spin operators, a useful expression of the concurrence has been obtained,\cite{WangZ2002}
\begin{eqnarray}
&&\mathcal{C}(\rho_{1N})
\nonumber \\
&&~= 2 \max\left[ 0, 
2|C^x_{1N}| - \sqrt{\left( \frac{1}{4} + C^z_{1N}\right)^2 - M^2} \right],
\nonumber \\
\label{eq:Conc_expect}
\end{eqnarray}
where $M = \langle S^z_1 \rangle = \langle S^z_N \rangle$ and $C^a_{1N} = \langle S^a_1 S^a_N \rangle$~($a=x,z$).
We calculate numerically the correlation functions $C^x_{1N}$ and $C^z_{1N}$ at zero and finite temperatures (note that $M=0$ in our model without external magnetic field), and then evaluate the concurrence using Eq.\ (\ref{eq:Conc_expect}).

\subsection{Ground state}\label{subsec:groundstate}

\begin{figure}
\begin{center}
\includegraphics[width=75mm]{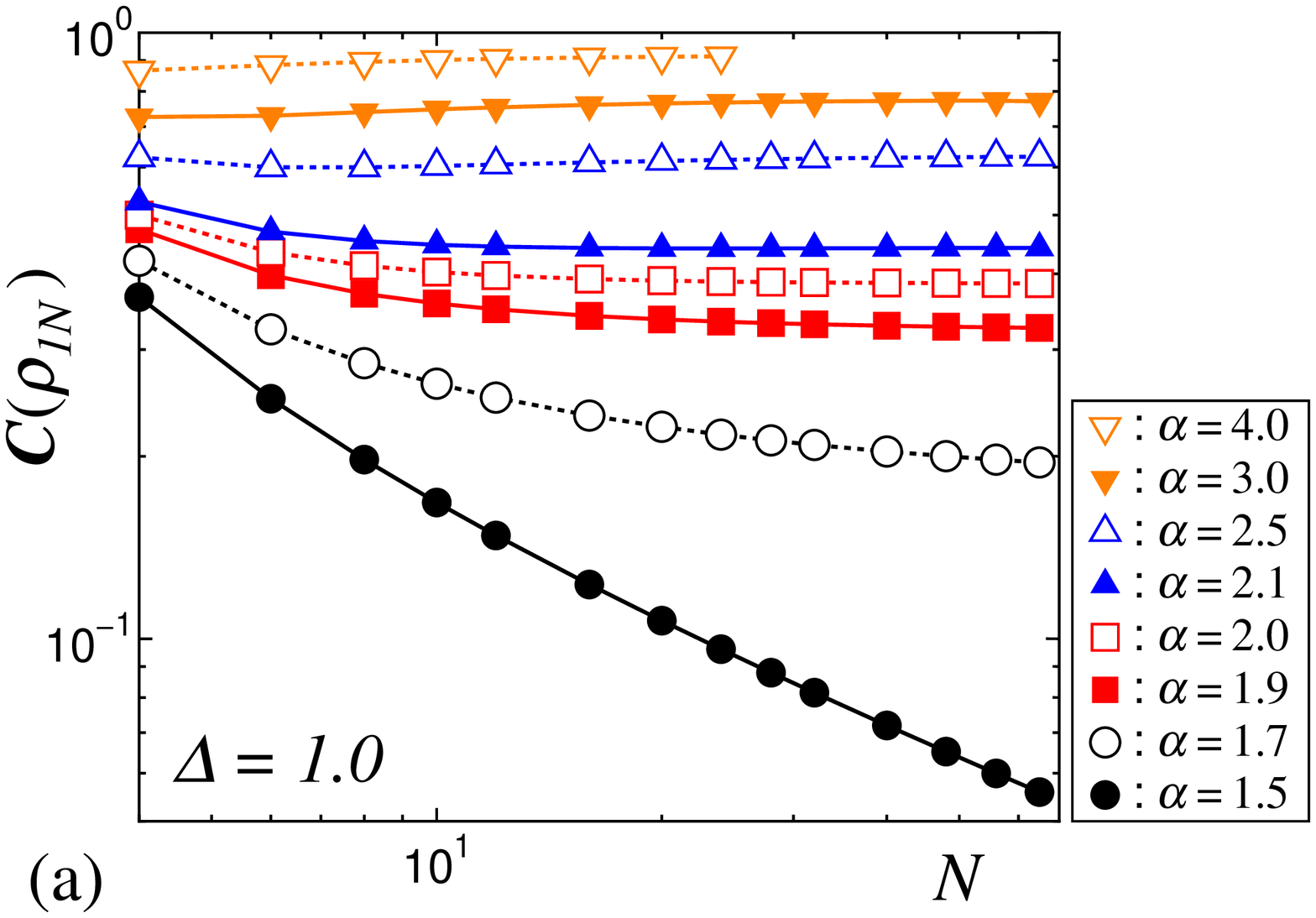}
\includegraphics[width=75mm]{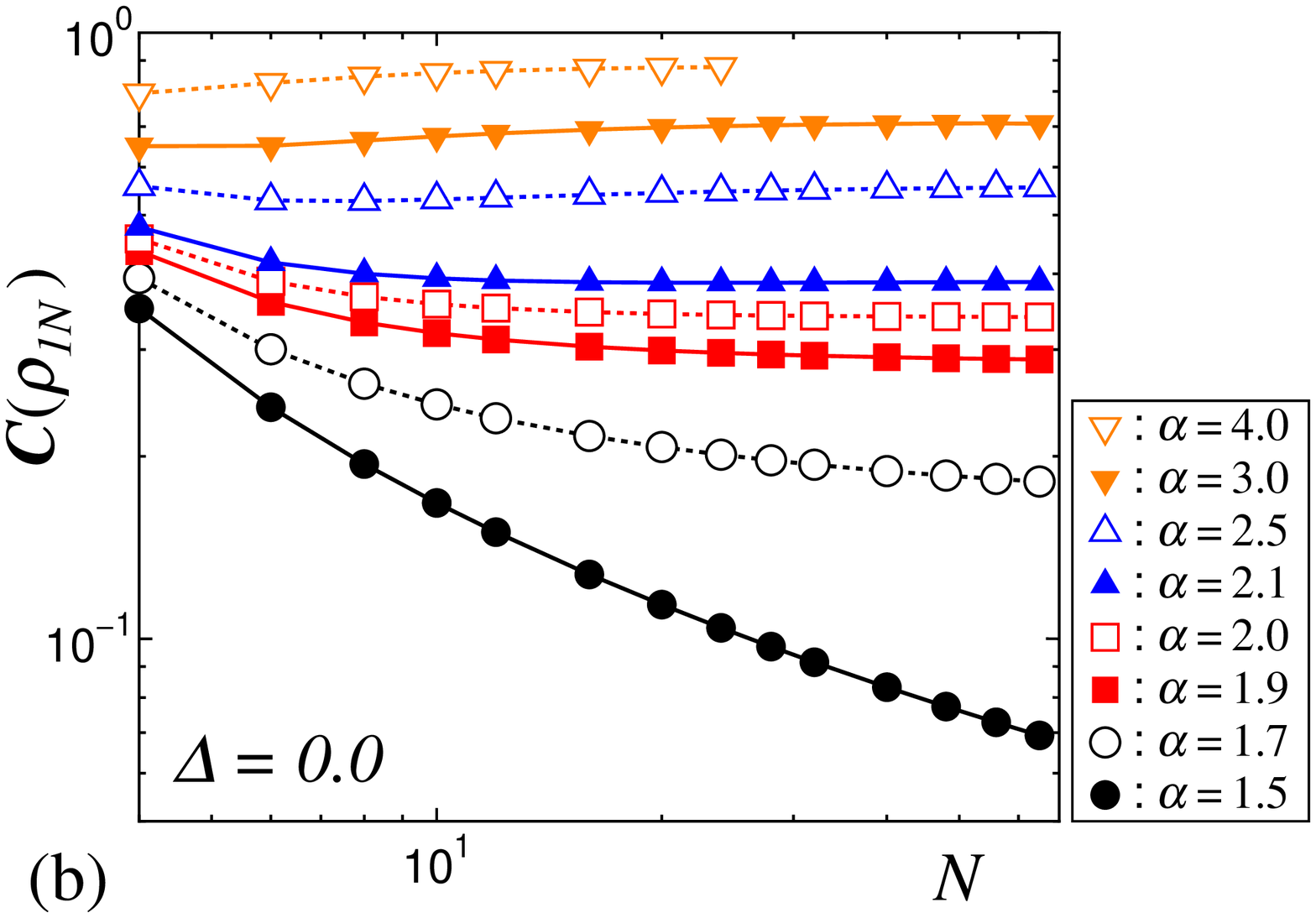}
\includegraphics[width=65mm]{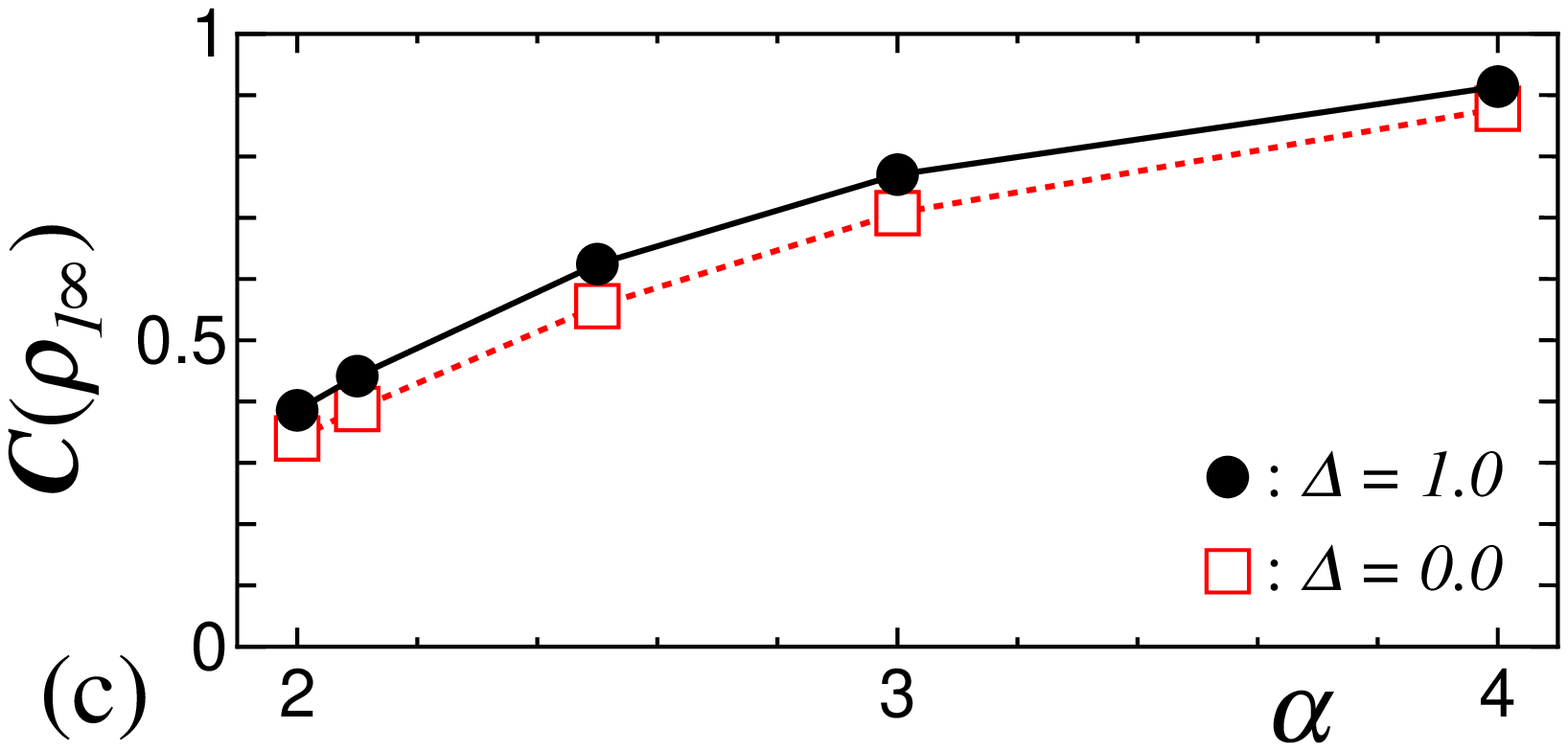}
\caption{(Color online) 
End-to-end concurrence $\mathcal{C}(\rho_{1N})$ in the ground state of the system with the sinusoidal deformation [Eq.\ (\ref{eq:Ham})] as functions of system size $N$; 
(a) Heisenberg case ($\Delta = 1$) and (b) XY case ($\Delta = 0$).
Estimates of $\mathcal{C}(\rho_{1N})$ at $N \to \infty$ are shown in (c) as functions of $\alpha$.
Lines are guides for the eye.
}
\label{fig:conc_gs}
\end{center}
\end{figure}

In Fig.\ \ref{fig:conc_gs}, we show our numerical data of the end-to-end concurrence in the ground state ($T=0$) of the SD systems with up to $N=64$ spins.
The data for $N \le 24$ were obtained by the exact-diagonalization method while those for $N \ge 28$ were calculated by the density-matrix renormalization group (DMRG) method.\cite{White1992,White1993}
Figures \ref{fig:conc_gs} (a) and (b) show the system-size dependence of the concurrence for the Heisenberg ($\Delta=1$) and XY ($\Delta=0$) cases, which exhibit essentially the same behavior.
It is clear that for $\alpha \ge 2$ the concurrence takes a finite value even at the thermodynamic limit $N \to \infty$.
For $\alpha=4$, we could not obtain the data for $N \ge 28$ since the DMRG calculation did not converge the true ground state due to the extremely small energy scale around open edges.
However, the exact-diagonalization results for $N \le 24$ are enough to suggest the convergence of the data to a finite large value at $N \to \infty$.
For $\alpha < 2$, on the other hand, the end-to-end concurrence decreases slowly, presumably in a power law, with the system size $N$.
We note that the concurrence for $\alpha=1.9$ decreases very slowly but monotonically with $N$.
From the results, we conclude that the system with $\alpha \ge 2$ realizes the LDE.

Figure\ \ref{fig:conc_gs}(c) shows the end-to-end concurrence in the thermodynamic limit for the LDE regime as a function of $\alpha$.
As estimates of the concurrence at $N \to \infty$, $\mathcal{C}(\rho_{1N})$ for $N=64$ ($N=24$) is plotted for $\alpha \le 3.0$ ($\alpha=4.0$).
The figure suggests that the end-to-end concurrence in the ground state is monotonically increasing with $\alpha$.
This is consistent with an intuitive argument that the correlation between an edge spin ${\bf S}_1$ (${\bf S}_N$) and the neighboring spin ${\bf S}_2$ (${\bf S}_{N-1}$) becomes weaker as the ratio of the edge bond to the bond next to the edge is smaller, which results in a larger entanglement between the edge spins from the monogamy condition.
We note that for $\alpha=2$ the end-to-end concurrence at $T=0$ is equivalent to the concurrence between the nearest-neighboring spins in a uniform periodic system as discussed in Sec.\ \ref{subsec:LDE_in_SSD}.
Indeed, the numerical results for $\alpha=2$ coincide with the exact value obtained from the nearest-neighbor spin correlations in the infinite uniform spin chain: $\lim_{N \to \infty} \mathcal{C}(\rho_{1N})=0.3863...$ for the Heisenberg chain ($\Delta=1$) and $\lim_{N \to \infty} \mathcal{C}(\rho_{1N})=0.3393...$ for the XY chain ($\Delta=0$).

\subsection{Finite temperatures}\label{subsec:finiteT}

\begin{figure}
\begin{center}
\includegraphics[trim=0 0 75 75, width=70mm]{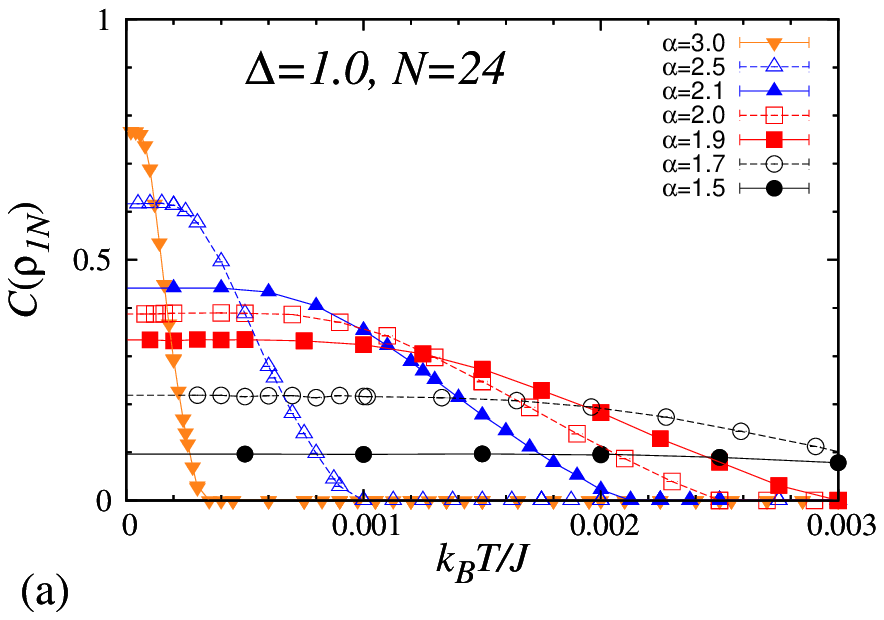}
\includegraphics[trim=0 0 75 75, width=70mm]{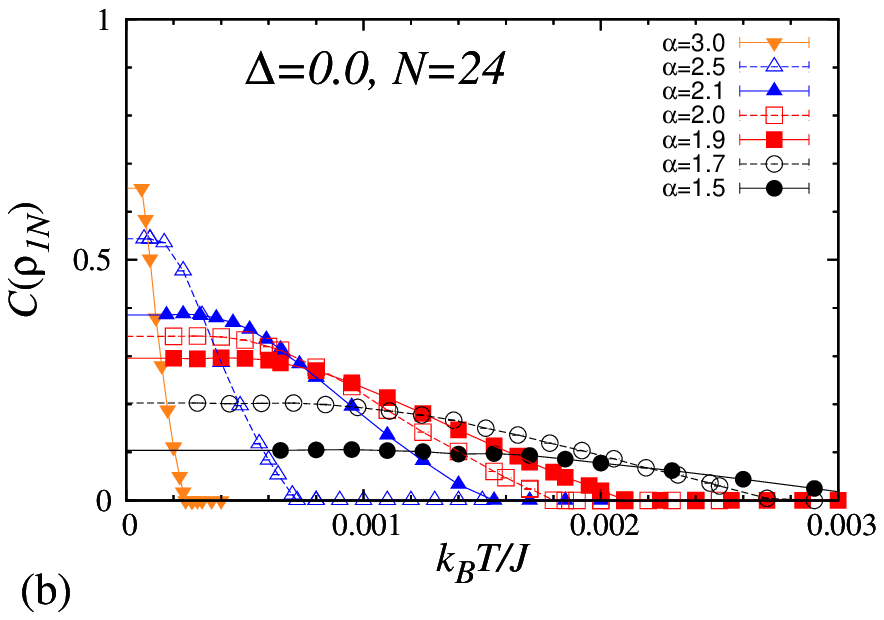}
\caption{(Color online) 
Temperature dependence of the end-to-end concurrence $\mathcal{C}(\rho_{1N})$ in the system under the sinusoidal deformation [Eq.\ (\ref{eq:Ham})] with $N=24$ spins; 
(a) Heisenberg case ($\Delta = 1$) and (b) XY case ($\Delta = 0$). 
All data are plotted with error bars but the most of them are smaller than the symbol size.
Lines are guides for the eye.
}
\label{fig:conc_finiteT}
\end{center}
\end{figure}

In the previous section, we have shown that the SD system with $\alpha \ge 2$ exhibits the LDE.
The end-to-end concurrence in the ground state increases with $\alpha$, suggesting that larger $\alpha$ is preferable for achieving a larger amount of LDE.
However, since the SD with large $\alpha$ leads to a small energy scale around the open edges, it is naturally expected that the system with large $\alpha$ becomes fragile against thermal fluctuations.
Therefore, it is important to examine how robustly the LDE in the system with SD survives at finite temperatures.
To the end, we have calculated the end-to-end concurrence at finite temperatures by the quantum Monte-Carlo method based on the directed-loop (worm) algorithm.\cite{Worm,DLA,OrderN}

In Fig.\ \ref{fig:conc_finiteT}, we present the numerical data of the end-to-end concurrence at finite temperatures for $N=24$ as a typical example.
For all the cases of $\alpha$ and $\Delta$ calculated, we have observed qualitatively the same behavior: 
When $T$ increases from zero, the concurrence remains at almost the same value as that for $T=0$ for some temperature range, and then decreases smoothly down to zero.
As $\alpha$ is larger, the concurrence is larger at $T \to 0$ but starts to decrease at a lower temperature, as expected.\cite{QMCalpha4}
On the other hand, when $\alpha$ is small, the concurrence is rather small at $T \to 0$ but can survive up to high temperatures.
Therefore, there is a trade-off between the amount of LDE achievable at $T=0$ and the temperature range for which the LDE is sustainable.
For a practical purpose, one may choose $\alpha$ to maximize the amount of LDE at temperatures achievable.
Our data may serve as a basis for the judgement.

\begin{figure}
\begin{center}
\includegraphics[trim=0 0 75 75, width=70mm]{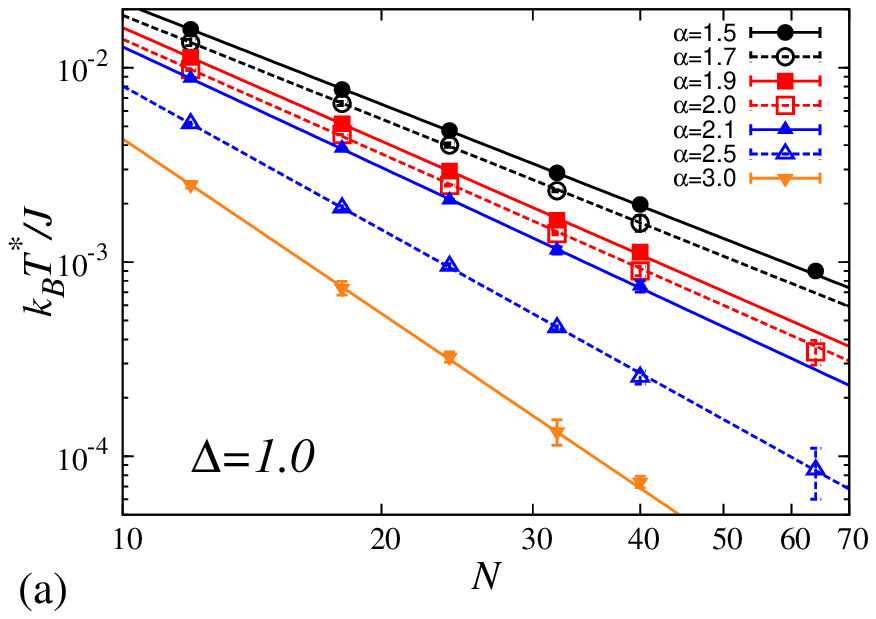}
\includegraphics[trim=0 0 75 75, width=70mm]{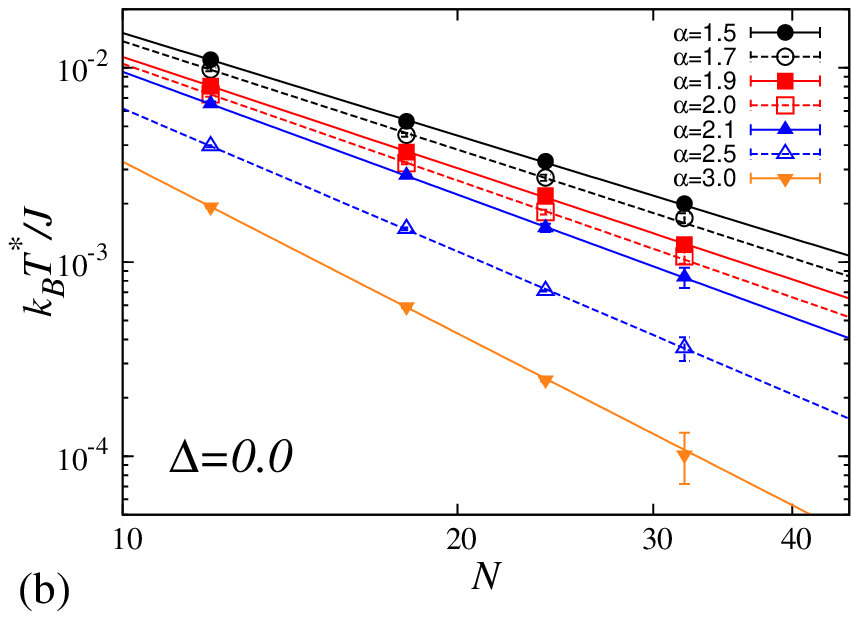}
\includegraphics[trim=0 0 75 75, width=70mm]{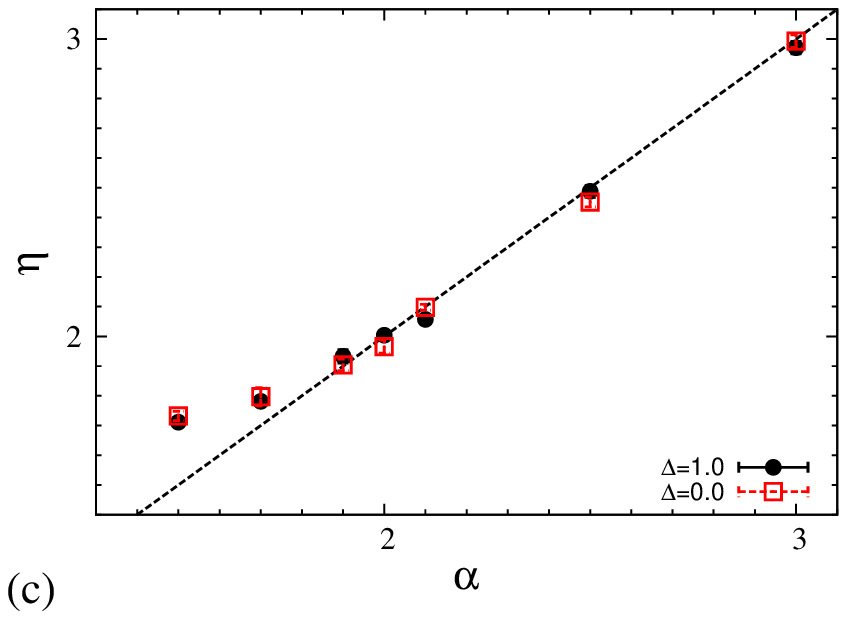}
\caption{(Color online) 
Temperature $T^*$ at which the end-to-end concurrence vanishes as functions of system size $N$; 
(a) Heisenberg case ($\Delta = 1$) and (b) XY case ($\Delta = 0$).
Lines show the fits to an algebraically decaying form $k_B T^*/J = A N^{-\eta}$ with fitting parameters $\eta$ and $A$.
(c) Decay exponent $\eta$ as functions of $\alpha$.
The dotted line represents the relation $\eta = \alpha$.
All data are plotted with error bars.
}
\label{fig:Tstar-N}
\end{center}
\end{figure}

To evaluate the robustness of LDE more quantitatively, we examine the temperature $T^*$ at which the end-to-end concurrence $\mathcal{C}(\rho_{1N})$ vanishes. 
Figures\ \ref{fig:Tstar-N} (a) and (b) show the system-size dependence of $T^*$.
The data clearly suggest that $T^*$ decreases with $N$ algebraically, $k_B T^*/J \sim N^{-\eta}$.
The decay exponent $\eta$ obtained from the fitting is plotted in Fig.\ \ref{fig:Tstar-N} (c) as a function of $\alpha$.
For $\alpha \ge 2$, the exponent obeys a relation $\eta = \alpha$.
This behavior is not trivial but can be understood from the fact that the energy scale at edge bonds, which should give the energy scale of the excited state localized around the edges, changes as $\sin^\alpha(\pi/N) \sim N^{-\alpha}$.
For $\alpha < 2$, on the other hand, the decay exponent $\eta$ is larger than $\alpha$.
While this numerical observation indicates the presence of low-energy excitations which has an excitation energy smaller than that of the edge states, it is not clear how to construct such an excited state.
This question for $\alpha < 2$ is open for future research.

\section{concluding remarks}\label{sec:Conclusion}

We have studied the long distance entanglement between edge spins of one-dimensional spin systems under the SD. 
In the systems, the energy scale of the local Hamiltonian is modified smoothly from a maximum at the center of the system to zero at open edges according to a rescaling function Eq.\ (\ref{eq:fx}).
When the exponent in the rescaling function is $\alpha=2$, it was shown that the ground state of the system was equivalent to the one of the corresponding uniform periodic system.
Therefore, the system under SD with $\alpha=2$ generates an entanglement between edge spins whose strength is the same as that between nearest-neighboring spins in the periodic chain.
The system thereby realizes the long-distance entanglement, a finite entanglement between edge spins infinitely far apart from each other, in the ground state.

We have investigated numerically the entanglement between the edge spins in the SD system for various $\alpha$.
Using the exact diagonalization and DMRG methods, we have calculated the end-to-end concurrence in the ground state.
It is then found that the SD with $\alpha \ge 2$ generates LDE.
The amount of the entanglement is larger as $\alpha$ is larger.
We have also examined how robust the LDE is against thermal fluctuations.
The numerical data of the end-to-end concurrence at finite temperatures obtained from the quantum Monte-Carlo method suggest that when temperature increases, the concurrence keeps the value at $T=0$ for some range of low temperatures, then disappears smoothly.
The temperature $T^*$ at which LDE vanishes decays with the system size $N$ in a power law, $k_B T^*/J \sim N^{-\eta}$.
For $\alpha \ge 2$, the decay exponent $\eta$ is found to be equal to $\alpha$.
Our numerical results thus indicate that large $\alpha$ is preferable for generating a large LDE at $T=0$ but disadvantageous in sustaining LDE at finite temperatures.
For a practical purpose, one may select the optimal $\alpha$ in the energy deformation considering the amount of LDE required and the lowest temperature which can be achieved.

A characteristic feature of SD is that it realizes a ``true" LDE in the ground state in the sense that the entanglement between edge spins remains finite even at the infinitely long chain.
Furthermore, the LDE can survive up to a rather high temperature; the temperature $T^*$ at which LDE vanishes decays quite slowly, in a power law, with the system size.
The latter point is in contrast with the LDE mediated by systems with gapful excitations, for which $T^*$ typically decays exponentially with system size.
We also emphasize that our results are not restricted to the spin-1/2 chains but applicable to general 1D critical systems belonging to the universality class of Tomonaga-Luttinger liquid.
These properties of SD would make it be of wide application in entanglement engineering using quantum correlated systems.

The realization of a system with SD in laboratory is an intriguing and challenging problem.
The systems of ultra-cold atoms in optical lattices, for which hopping amplitudes and interatomic couplings can be tuned\cite{Roati2008,Bakr2009,Yamazaki2010}, should be a promising candidate.

\acknowledgments
The authors thank to Hosho Katsura, Isao Maruyama, Konstantin Matveev, Tomoyuki Morimae, and Tomotoshi Nishino for fruitful discussions.
T.H. was supported by Grants-in-Aid for Scientific Research from MEXT, Japan (Grant No. 24740255). 
Numerical calculations were performed in part at the ISSP Supercomputer Center of the University of Tokyo and cluster machines in Research center for Nano-micro structure science and engineering, University of Hyogo.


\begin{thebibliography}{99}
\bibitem{QMCalpha4}
We note that for $\alpha=4$ and $N=24$, we have performed the quantum Monte-Carlo calculation for temperatures down to $T=6 \times 10^{-5} J / k_B$ and found that $\mathcal{C}(\rho_{1N})=0$ for the temperature range.
We expect that the end-to-end concurrence rises at a lower temperature and converges the value at $T=0$ shown in Figs.\ \ref{fig:conc_gs} (a) and (b) as temperature decreases.
\end{thebibliography}
\end{document}